\documentclass[twocolumn]{aastex631}

\shorttitle{Asymmetry in the Narrow Fe K$\alpha$ Line in MCG-5-23-16}
\shortauthors{Victor Liu, Abderahmen Zoghbi, and Jon M. Miller}

\usepackage{graphics}
\graphicspath{{./}{figures/}}

\usepackage{soul}
\usepackage{multirow}
\usepackage{orcidlink}
\usepackage{microtype}

\edef\restoreparindent{\parindent=\the\parindent\relax}
\usepackage{parskip}
\restoreparindent

\begin{document}

\title{Detection of Asymmetry in the Narrow Fe K$\alpha$ Emission Line in MCG-5-23-16 with \textit{Chandra}}

\correspondingauthor{Victor Liu}
\email{v.liu@yale.edu}

\author{Victor Liu \orcidlink{0000-0003-4434-3921}}
\affiliation{Department of Astronomy, Yale University, 219 Prospect St, New Haven, CT 06511, USA}
\affiliation{HEASARC, Code 6601, NASA/GSFC, Greenbelt, MD 20771, USA}
\affiliation{CRESST II, NASA Goddard Space Flight Center, Greenbelt, MD 20771, USA}

\author{Abderahmen Zoghbi \orcidlink{0000-0002-0572-9613}}
\affiliation{HEASARC, Code 6601, NASA/GSFC, Greenbelt, MD 20771, USA}
\affiliation{CRESST II, NASA Goddard Space Flight Center, Greenbelt, MD 20771, USA}
\affiliation{Department of Astronomy, University of Maryland, College Park, MD 20742, USA}

\author{Jon M. Miller \orcidlink{0000-0003-2869-7682}}
\affiliation{Department of Astronomy, University of Michigan, 1085 South University Ave, Ann Arbor, MI 48103, USA}

\begin{abstract}
Iron K$\alpha$ (Fe K$\alpha$) emission is observed ubiquitously in AGN, and it is a powerful probe of their circumnuclear environment. Examinations of the emission line play a pivotal role in understanding the disk geometry surrounding the black hole. It has been suggested that the torus and the broad line region (BLR) are the origins of emission. However, there is no universal location for the emitting region relative to the BLR. Here, we present an analysis of the narrow component of the Fe K$\alpha$ line in the Seyfert AGN MCG-5-23-16, one of the brightest AGN in X-rays and in Fe K$\alpha$ emission, to localize the emitting region. Spectra derived from \textit{Chandra/HETGS} observations show asymmetry in the narrow Fe K$\alpha$ line, which has only been confirmed before in the AGN NGC 4151. Models including relativistic Doppler broadening and gravitational redshifts are preferred over simple Gaussians and measure radii consistent with $R \simeq$ 200-650 \textit{r$_g$}. These results are consistent with those of NGC 4151 and indicate that the narrow Fe K$\alpha$ line in MCG-5-23-16 is primarily excited in the innermost part of the optical broad line region (BLR), or X-ray BLR. Characterizing the properties of the narrow Fe K$\alpha$ line is essential for studying the disk geometries of the AGN population and mapping their innermost regions.

\end{abstract}

\section{Introduction} \label{sec:intro}
Iron K$\alpha$ emission lines (Fe K$\alpha$) at 6.4 keV are observed ubiquitously in AGN \citep{nandra_pounds1994, weaver2001, shu2010agnsurvey}. The line is produced in dense and cold material when illuminated by an X-ray source, making it a powerful probe of circumnuclear environments in AGN. A significant portion of the information we can learn about the inner regions of AGNs and their geometries comes from modeling the line and understanding what emission processes may be responsible for it.

However, the location where the Fe K$\alpha$ line is emitted is not well established. It is thought that the dust sublimation radius forms an outer envelope to the emitting region \citep{netzer_laor1993, czerny_hryniewicz2011, gandhi2015, baskin_laor2017}. Although \citet{nandra2006} found no correlation between the width of the Fe K$\alpha$ and the optical H$\beta$ line in a sample of sources, implying that the Fe K$\alpha$ line originates in the torus and not the optical BLR, other observations suggest otherwise (e.g \citet{bianchi2008}). \citet{shu2010agnsurvey} presented an extensive analysis of the Chandra grating spectra of 36 sources, ﬁnding that there is no universal location for the Fe K$\alpha$ line-emitting region relative to the optical BLR. They found that the Fe K$\alpha$ line may have contributions from parsec-scale distances from the black hole (i.e. the torus), down to matter on the optical BLR scale or smaller. 

Direct measurements of the location of the line emitting regions was recently obtained for NGC 4151, the brightest Seyfert galaxy in X-rays, and hence the AGN with the brightest narrow Fe K$\alpha$ line. Using Chandra grating spectra, \citet{miller2018_NGC4151} found that the narrow Fe K$\alpha$ line in NGC 4151 is asymmetric, with the data suggesting that the emission could originate in a region as small as 50–130 r$_g$ during high ﬂux intervals\footnote{The gravitational radius r$_g$ is a characteristic length scale associated with black holes and represents the radius at which the gravitational pull of a black hole becomes significant compared to other forces. The formula for calculating r$_g$ is: $r_g = GM/c^2$.}. On a parallel front, \citet{zoghbi2019_NGC4151} measured the time delay between the narrow Fe K$\alpha$ line and the hard X-ray continuum in NGC 4151 to constrain the location of the narrow Fe K$\alpha$ region--a technique called reverberation mapping--and found a time delay of 3.4$^{+2.5}_{-0.8}$ days. \citet{bentz2006} conducted a similar reverberation mapping study 13 years prior, but of the optical BLR in NGC 4151, and found a time delay of 6.6$^{+1.1}_{-0.8}$ days between the optical H$\beta$ line and the continuum. These results indicate that the narrow Fe K$\alpha$ line in NGC 4151 is primarily excited in either the innermost part of the optical broad line region (BLR), the X-ray BLR, or closer. The central question is now whether other AGN exhibit asymmetry and reverberation in their narrow Fe K$\alpha$ lines, which will crucially inform us about where the line is emitted more broadly in the population of all AGN.

Before proceeding further, we wish to make a distinction between the narrow Fe K$\alpha$ line and the broad Fe K$\alpha$ line. The narrow component of the line originates at large distances from the black hole ($\sim$ 100's to 1000's of \textit{r$_g$}) while the broad component originates less than 10 \textit{r$_g$} in the inner disk \citep{zoghbi2013unevenTimeLags, uttley2014reverberation}. While asymmetry and reverberation delays have been seen before in several sources' broad Fe K$\alpha$ lines, in this study, we are focusing on only the narrow Fe K$\alpha$ line.

Among the next brightest sources in the Fe K$\alpha$ band is MCG-5-23-16 (\textit{z} = 0.00849). MCG-5-23-16 is a Seyfert 1.9-AGN \citep{veron1980galaxies} with a central black hole mass of 10$^{7.9} M_\sun$ \citep{ponti2013}. There have been many studies of the narrow and broad components of the Fe K$\alpha$ line in MCG-5-23-16 (e.g. \citet{reeves2006, reeves2007suzaku, zoghbi2014, zoghbi2017mcg_nustar}), but none using the high-energy resolution afforded by \textit{Chandra}, which is crucial in order to accurately model the narrow Fe K$\alpha$ line. In this paper, we use 12 \textit{Chandra} observations of MCG-5-23-16, nine of which were recently taken in 2020, to model the narrow Fe K$\alpha$ line and address the question of whether MCG-5-23-16 also has asymmetry in its narrow Fe K$\alpha$ line like NGC 4151. We find that MCG-5-23-16 does indeed have line asymmetry, and that the strength of this line asymmetry varies with time.

Here, we report on the narrow Fe K$\alpha$ line and its model fit parameters in MCG-5-23-16 using observations with \textit{Chandra}.

Our specific aims are: 
\begin{enumerate}
    \item To determine whether the narrow Fe K$\alpha$ line in MCG-5-23-16 exhibits asymmetry, just like it does in NGC 4151.
    \item To examine any potential variability in the narrow Fe K$\alpha$ line.
    \item To constrain the emission processes at play in the disk and the geometry of the AGN by trying out different models for the line and examining their best-fit parameters.
\end{enumerate}

\section{Observations and Reduction} \label{sec:reduction}

All archival Chandra/\textit{HETG} observations of MCG-5-23-16 were downloaded directly from the \textit{Chandra} archive. In addition to these archival observations, we observed MCG-5-23-16 nine times from October to November 2020 with the \textit{Chandra X-ray Observatory, Chandra} \citep{weisskopf2000chandra} using the Advanced CCD Imaging Spectrometer (ACIS) optimized for spectroscopy (ACIS-S) in faint mode using the High Energy Grating Transmission Spectrometer (HETGS). The full dataset used in this paper is available for download at \dataset[DOI: 10.25574/cdc.186]{https://doi.org/10.25574/cdc.186}. The observation identification number (ObsID), start date, and duration of each exposure is given in Table \ref{tab:observations}.

HETGS observations typically provide data from both the high-energy gratings (HEG) and the medium-energy gratings (MEG). However, the MEG has less effective area in the Fe K band and lower resolution, and is therefore less suited to our analysis. As a result, we limited our analysis to only the HEG. 

The standard CIAO tools (version 4.15.1) were used to reduce the \textit{Chandra} observations \citep{fruscione2006ciao}. For each observation, with the exception of ObsID 2121, we ran the tool ``chandra\_repro'' to produce the necessary ``evt2'', ``pha2'', RMF and ARF files. For ObsID 2121, ``chandra\_repro'' produced evt2 and pha2 files with erroneous exposure times, so we manually generated the spectrum and filtered for bad grades and for a "clean" status column using the tools ``tgdetect2'', ``tg\_create\_mask'', ``tg\_resolve\_events'', ``dmcopy'', ``dmappend'', and ``tgextract'' as according to the CIAO HETG/ACIS-S Grating Spectra thread.\footnote{\href{https://cxc.cfa.harvard.edu/ciao/threads/spectra_hetgacis/}{\url{https://cxc.cfa.harvard.edu/ciao/threads/spectra_hetgacis/}}} The RMF and ARF files were then generated using ``mktgresp''.

For each exposure, we added the first-order (+1 and -1) HEG spectra, RMF files, and ARF files using ``combine\_grating\_spectra''. We found that individual spectra lacked sufficient data to constrain spectral features and fit complex physical models. When fit to a continuum using the model \textit{zphabs*zpowerlw}, individual spectra had a consistent continuum, with each spectra having a photon index value between 1.64 and 1.70. Therefore, we further combined the spectra using ``combine\_spectra'' into a ``Total'' spectrum, ``2000,2005'' spectrum and ``2020'' spectrum, corresponding to all 12 observations, the three observations in 2000 and 2005, and the nine observations taken recently in 2020 respectively. This grouping by time allows us to investigate how the Fe K$\alpha$ line in MCG-5-23-16 has changed from the earlier observations in 2000 and 2005 to the recent observations in 2020. From henceforth, ``the spectra'' will refer to the ``Total'', ``2000,2005'', and ``2020'' groups.

The spectra were then binned to achieve a minimum signal-to-noise ratio (SNR) of 5.0 for each energy bin using the tool ``ftgrouppha'' within the standard HEASFOFT suite, version 6.31.1. We set the ``grouptype'' parameter within ``ftgrouppha'' to be ``optsnmin'', which uses the optimal binning algorithm designed by \citet{kaastra2016optimal}.

All spectra fits were performed using PyXspec, version 2.1.2 \citep{craig2021pyxspec}, on the HEAsoft environment \citep{heasarc2014}. Unless otherwise noted, the errors quoted in this work reﬂect the value of the parameter at its 1$\sigma$ confidence interval. Errors were derived using the standard PyXspec “Fit.error()” command. The script used to perform the spectral fits and generate Figures \ref{fig:gaussian_fits_by_time} and \ref{fig:reflection_fits_by_time} in this paper is available on GitHub\footnote{\url{https://github.com/victorliu1231/mcg_5_23_16_chandra_script}} and version v1 is archived in Zenodo \citep{liu2023scriptForPlots}.

\begin{deluxetable}{l|l|l}[t]
    \tablecaption{\textit{Chandra} Observations of MCG-5-23-16 \label{tab:observations}}
    \tablehead{\colhead{ObsID} & \colhead{Date} & \colhead{Exp. (ks)}}
    \startdata
        2121 & 2000 Nov. 14 & 76.22 \\
        6187 & 2005 Dec. 08 & 30.08 \\ 
        7240 & 2005 Dec. 09 & 20.25 \\ 
        22553 & 2020 Oct. 11 & 16.30 \\ 
        24753 & 2020 Oct. 11 & 42.08 \\ 
        22554 & 2020 Oct. 17 & 29.08 \\ 
        24833 & 2020 Oct. 18 & 27.08 \\ 
        22555 & 2020 Oct. 26 & 25.23 \\ 
        24849 & 2020 Oct. 28 & 30.08 \\ 
        22556 & 2020 Nov. 17 & 20.08 \\ 
        24863 & 2020 Nov. 18 & 15.08 \\ 
        24864 & 2020 Nov. 20 & 25.08 \\ 
    \enddata
    \tablecomments{The list of \textit{Chandra} observations that we used for MCG-5-23-16. Exp. denotes the total exposure time for each observation in ks (kiloseconds). The total exposure time is 356.64 ks.}
\end{deluxetable}

\section{Spectral Modeling} \label{sec:modeling}
\subsection{Spectral Fitting Range and Setup} \label{subsec:spectra_setup}
The spectra in this work are restricted to the 1.5-8.0 keV range due to the low SNR below and above these energies. Additional bad data points were ignored with the command ``AllData.ignore(`bad')''.

We fit the spectra using three Gaussian models and three reflection models. Of the Gaussian models, the first model is \textbf{Model A: \textit{zphabs*(zpowerlw+zgauss)} with fixed $\sigma$=0}, a narrow Gaussian model that probes whether the Fe K$\alpha$ line originates at the torus or outer BLR and tests whether the line is resolved by \textit{Chandra}/HETGS. The next model is \textbf{Model B: \textit{zphabs*(zpowerlw+zgauss)} with free $\sigma$}, a broad Gaussian model that models emission from the intermediate regions of the AGN slightly closer to the central black hole. These first two models are both simple Gaussians and assume the Fe K$\alpha$ line is symmetric. The next models we use are more physically motivated and account for potential asymmetry in the line. We next try \textbf{Model C: \textit{zphabs*(zpowerlw+rdblur*zgauss)}}, which fits to asymmetry in the line by taking into account relativistic Doppler broadening near the black hole with the ``rdblur'' component (``rdblur'' is a convolution model extracted from the diskline model described in \citet{fabian1989diskline}).

We also attempt variants of Models A, B, and C, where instead of fitting a single Gaussian to the Fe K$\alpha$ line, we fit a double Gaussian. Fitting a double Gaussian accounts for the fact that the Fe K$\alpha$ has two line energies, one at 6.404 keV (Fe K$\alpha$1) and another at 6.391 keV (Fe K$\alpha$2). However, when we attempt these double Gaussian models, we find that the fits worsen compared to the single Gaussian models, likely due to the difference between these two line energies being unresolved in our spectra.

Afterwards, we try reflection models. We first attempt \textbf{Model D: \textit{zphabs*(zpowerlw+xillver)}}, which fits to asymmetry in the Fe K$\alpha$ line by taking into account Compton scattering of line photons reflecting off the disk with the ``xillver'' model \citep{garcia2010xillver1, garcia2013xillver3}. The specific table we use here is ``xillver-a-Ec5.fits''. Next, we use \textbf{Model E: \textit{zphabs*(zpowerlw+mytorus)}} using the table ``mytl\_V000010nEp000H500\_v00.fits'' \citep{murphy_yaqoob2009mytorus}. The ``mytorus'' component models Compton-scattering off of a toroidal structure around the disk. Finally, we use \textbf{Model F: \textit{zphabs*(zpowerlw+rdblur*mytorus)}} to determine whether including relativistic Doppler broadening makes a difference to the fit.

For all models, we obtained the Fe K$\alpha$ line flux by multiplying the component \textit{cflux} by the last additive component (e.g. \textit{cflux*zgauss, cflux*rdblur*zgauss}, \textit{ cflux*xillver, cflux*mytorus} \textit{and cflux*rdblur*mytorus}). The \textit{E$_{min}$} and \textit{E$_{max}$} parameters for \textit{cflux} were set to 6.2 and 6.45 keV, respectively, which generally bounded the entirety of the Fe K$\alpha$ emission line. In all fits, the emission line energy was frozen at 6.4 keV, the intrinsic line energy. Within Model C, $\sigma$ was fixed to 0 since the \textit{R$_{in}$} parameter from \textit{rdblur} was responsible for capturing the width of the line instead of \textit{zgauss}, and the outer radius of the disk \textit{R$_{out}$} was fixed at 10$^6$ \textit{r$_g$} to give the biggest range for \textit{R$_{in}$} to vary. When fitting to the ``2000,2005'' spectrum, letting inclination range from 0$^\circ \leq i \leq$ 90$^\circ$ produced a physically implausible lower bound of less than 0$^\circ$, so we restricted the inclination to range from 3$^\circ \leq i \leq$ 87$^\circ$ for the ``2000,2005'' spectrum. 

Within Model D, the photon index of \textit{zpowerlw} and \textit{xillver} were linked to get a consistent photon index for the continuum; the ionization index \textit{log$\xi$} was fixed at 0 since the Fe K$\alpha$ line is a neutral line; and the high energy cutoff \textit{E$_{cut}$} was fixed to 300 keV. In Model E and Model F, the hydrogen column density \textit{n$_H$} of \textit{mytorus} was tied to the \textit{n$_H$} of \textit{zphabs}, and the photon index of \textit{mytorus} was linked to the photon index of \textit{zpowerlw}. In Model E, the inclination was fixed to 60$^\circ$ and in Model F, the inclination and line emissivity was free to vary.

\subsection{Gaussian Models} \label{subsec:gaussianModels}
\subsubsection{Model A} \label{subsubsec:modelA}
Model A is zphabs*\textit{(zpowerlw+zgauss)} with $\sigma$ fixed to 0. In addition to testing whether the Fe K$\alpha$ line emitting region originates at the torus or outer BLR, fixing sigma to 0 also examines whether the line width is below the instrumental resolution of \textit{Chandra}/HETGS. Fixing $\sigma$ = 0 eV assumes that the line is not resolved and that its width is due to instrumental broadening and not any physical effects from the AGN itself. Figure \ref{fig:gaussian_fits_by_time} shows that the line is clearly broader than the fit for the ``2000,2005'', ``2020'', and ``Total'' spectra, proving that the line is resolved. It also shows that the origin for the line is not located in the torus or outer BLR. These fits also illustrate the asymmetry of the Fe K$\alpha$ line in MCG-5-23-16, as there is a clear red wing.

\subsubsection{Model B} \label{subsubsec:modelB}
Model B is \textit{zphabs*(zpowerlw+zgauss)} with free $\sigma$. In Model B, we let $\sigma$ vary in order to fit a broad Gaussian to the line. For the ``2000,2005'' spectrum, this only marginally improves the fits, as the red wing of the line is still not captured (Figure \ref{fig:gaussian_fits_by_time}). For the ``2020'' and ``Total'' spectra, however, letting $\sigma$ vary vastly improves the fits, since most of the red wing is now captured. These results are supported by the F-tests in Table \ref{tab:ftest_by_time}: the change from Model A to Model B for the ``2000,2005'' spectrum has a 1.95$\sigma$ level of confidence, and a 5.33$\sigma$ and 5.41$\sigma$ level of confidence for the ``2020'' and ``Total'' spectra, respectively. 

In the ``2020'' and ``Total'' spectra, the width is consistent with $\sigma \simeq$ 22 eV, corresponding to a FWHM of 52 eV and a \textit{projected} velocity of v $\simeq$ 2427 km/s. These values are very similar to those obtained by \citet{miller2018_NGC4151} for NGC 4151 -- the width and FWHM for the broad Gaussian fits in their work is 23 eV and 54 eV, respectively, emphasizing the similarity between the Fe K$\alpha$ line in MCG-5-23-16 and in NGC 4151.

As stated in \citet{miller2018_NGC4151}, it's important to emphasize that the neutral iron K$\alpha$ line is actually a composite of two distinct lines at laboratory energies of 6.391 keV and 6.404 keV \citep{bambynek1972labEnergies}. The difference between these lab energies constitutes just 25\% of the measured FWHM = 52 eV, so it is unnecessary to model for both lines and modeling just a single broad Gaussian is sufficient. Noteworthy is the fact that the separation between these lines falls also below the predicted resolution of the first-order High Energy Grating (HEG) in the iron K band, which is around 45 eV, making modeling both lines infeasible. This decision is also followed in previous work on the narrow Fe K$\alpha$ line in AGN (e.g. \citet{shu2010agnsurvey} and \citet{miller2018_NGC4151}).

\subsubsection{Model C} \label{subsubsec:modelC}
Model C is \textit{zphabs*(zpowerlw+rdblur*zgauss)}. In Model C, we multiply \textit{zgauss} with the \textit{rdblur} convolution model, which introduces a relativistic Doppler broadening factor to the Gaussian \citep{fabian1989diskline}. This broadening factor creates a red tail in the Gaussian and thus fits to the asymmetry in the line. We first try fixing the line emissitivity at \textit{q} = 3 (Model C.1), which models an isotropically radiative and flat accretion disk.

We find that the fit for the ``2000,2005'' spectrum significantly improves when using Model C.1. Visually, the red wing of the Fe K$\alpha$ line is better captured by Model C.1 than all the previous models (top panel of Fig. \ref{fig:gaussian_fits_by_time}) and the F-test between Model B and Model C.1 indicates an improvement in the fit at the 2.74$\sigma$ level of confidence (Table \ref{tab:fits_by_time}). With this evidence, we take Model C.1 as the benchmark fit for the ``2000,2005'' spectrum. The fit for ``2020'' also improves when using Model C.1: visually, the red wing is captured fully by Model C.1 (middle panel of Fig. \ref{fig:gaussian_fits_by_time}). However, since the amplitude of the red wing is much smaller in the ``2020'' spectrum, the fit with Model C.1 is not significantly better than the fit with Model B. More convincingly, the F-test between Model B and Model C.1 gives only a 1.97$\sigma$ level of confidence (Table \ref{tab:fits_by_time}), statistically indicating that Model B sufficiently fits the line and that the asymmetry in the line for the ``2020'' spectrum is weak. The ``2020'' spectrum also gives much larger uncertainties in the inclination and inner radius R$_{in}$ than the ``2000,2005'' spectrum, further indicating that the asymmetry in the line for ``2020'' is much weaker than in the line for ``2000,2005''. The collected counts of the 2020 spectrum also are approximately 1.3 times higher than those in the 2000-2005 spectrum, lending further support to the weaker asymmetry of the Fe K$\alpha$ line in ``2020'' compared to ``2000,2005''. Therefore, we consider Model B instead of Model C.1 as the benchmark fit for the ``2020'' spectrum since further fits do not significantly improve the fit.

Most importantly, the F-test between the Model B and Model C.1 for the ``Total'' spectrum gives a 3.05$\sigma$ level of confidence (Table \ref{tab:ftest_by_time}), indicating that when the data from the observations in 2020 is added to the ``2000,2005'' group, there is just enough statistical significance to reach the standard minimum 3$\sigma$ level of confidence to support a conclusion. Based off of the 3.05$\sigma$ level of confidence, we can conclude that the Fe K$\alpha$ line in MCG-5-23-16 as a whole is asymmetric.

The difference between the Model B to Model C.1 F-tests and the best-fit inner disk radius $R_{in}$ values in the ``2000,2005'' group and the ``2020'' group demonstrate that the Fe K$\alpha$ region in this source is variable, and that the emitting region has moved farther from the center of the AGN between 2000+2005 and 2020.

Afterwards, we also try letting the line emissitivity q vary within the range 2.0 $\leq$ \textit{q} $\leq$ 4.0 (Model D.2). Comparing the reduced $\chi^2$ results listed in Table \ref{tab:fits_by_time}, only marginally better fits are achieved when the emissivity varies for all spectra. In fact, evolving from Model B to Model C.2 is slightly worse than evolving from Model B to Model C.1, as seen by how the F-test values are slightly lower in the evolution from Model B to Model C.2 as compared to Model B to Model C.1 (Table \ref{tab:ftest_by_time}). Relative to Model C.1, where \textit{q} is fixed, smaller radii are found, but the small improvement in $\chi^2/\upsilon$ signals that smaller radii are not required. Therefore, we take Model C.1 as our benchmark fit, and all conclusions made about the inclination and inner disk radius \textit{R$_{in}$} will reference the values in Model C.1 as seen in Table \ref{tab:fits_by_time}.

\subsection{Reflection Models} \label{subsec:reflectionModels}
\subsubsection{Model D} \label{subsubsec:modelD}
Model D is \textit{zphabs*(zpowerlw+xillver)}. Model D replaces the \textit{zgauss} component of the previous three models with the reflection model \textit{xillver} \citep{garcia2010xillver1,garcia2013xillver3}. From Figure \ref{fig:reflection_fits_by_time}, it is clear there is asymmetry in the line in the ``2000,2005'' spectrum. \textit{xillver} models the asymmetry in the line by taking into account Compton scattering of line photons due to reflection off the disk. The \textit{xillver} model parameters include the photon index of the illuminating power law spectrum, the iron abundance, the ionization of the disk, the high-energy cutoff of the power law, the inclination at which the emitting region is viewed, and the redshift of the source. The specific table used is ``xillver-a-Ec5.fits'', the ionization log$\xi$ is frozen to 0, and the high-energy cutoff for the power law E$_{cut}$ is frozen to 300 keV.

When the line was fitted to Model D, the fits worsened substantially compared to the Gaussian models (see Figure \ref{fig:gaussian_fits_by_time} and Figure \ref{fig:reflection_fits_by_time}). The reduced $\chi^2$ values increased across all three spectra compared to Model B and is even higher than the reduced $\chi^2$ value for Model A in most of the spectra (Table \ref{tab:fits_by_time} and Table \ref{tab:fits_by_time}). Because Model D fits the data so poorly, using a F-test to compare it to Models A and B is meaningless, so we did not Model D in our comparisons of models in Table \ref{tab:ftest_by_time}. This also means that the parameter values in Table \ref{tab:fits_by_time} for Model D should not be quoted as physically real.

The disk reflection model \textit{pexmon} gave a similarly bad fit \citep{nandra2007pexmon}. Therefore, we can conclude that the line asymmetry in MCG-5-23-16 line cannot be from a disk with only intrinsic reflection broadening.

\subsubsection{Model E} \label{subsubsec:modelE}
Model E is \textit{zphabs*(zpowerlw+mytorus)}. Model E uses the toroidal reflection model \textit{mytorus} \citep{murphy_yaqoob2009mytorus} instead of the disk reflection models \textit{xillver and pexmon}. When fitting to the narrow Fe K$\alpha$ line, we found that the fit is insensitive to the inclination angle. Therefore, for each spectrum, we fixed the inclination of Model E at 60$^\circ$. For consistent hydrogen column density and photon index values, we also tied the \textit{nH} and \textit{PhoIndx} parameters of the \textit{mytorus} component to those of the \textit{zphabs} and \textit{zpowerlw} components.

We found that the Model E does not capture the red wing of the Fe K$\alpha$ line in all three spectra and also doesn't capture the full blue wing in the ``2020'' and ``Total'' spectra (Fig. \ref{fig:reflection_fits_by_time}). The reduced $\chi^2$ values for Model E are also higher than those of both Model C.1 and Model C.2 (Table \ref{tab:fits_by_time}), indicating that Model E is a worse fit than Model C. This demonstrates that the relativistic Doppler broadening introduced by \textit{rdblur} is responsible for the asymmetry we observe in the narrow Fe K$\alpha$ line, and not the Compton-scattering off of a toroidal structure as modeled by \textit{mytorus}.

We did not include the F-tests involving Model E in Table \ref{tab:ftest_by_time} because the number of free parameters in Model E is less than or equal to the number of free parameters in all other models, and the F-test is used only when a new model has more free parameters than previous models (e.g. when the new model constitutes an increase in complexity).

\subsubsection{Model F} \label{subsubsec:modelF}
Model F is \textit{zphabs*(zpowerlw+rdblur*mytorus)}. Since introducing relativistic Doppler broadening improved the fits for the Gaussian models (from Model B to Model C), we convolved the \textit{mytorus} component by \textit{rdblur} in Model F to see if the Doppler broadening will similarly improve the \textit{mytorus} fits. Unlike the Model E fit, the Model F fit was not insensitive to the inclination so we let the inclination freely vary in Model F. We also tried both fixing the line emissivity \textit{q} to 3.0 and letting \textit{q} freely vary, and found that when \textit{q} was able to freely vary, the fit was better able to capture the red wing. Similarly to Model F, we tied the \textit{nH} and \textit{PhoIndx} parameters of the \textit{mytorus} component to those of the \textit{zphabs} and \textit{zpowerlw} components.

We found that adding the \textit{rdblur} component to the \textit{mytorus} model in Model F significantly improved the fit to the asymmetry in the Fe K$\alpha$ line compared to Model E (see Fig. \ref{fig:reflection_fits_by_time}). This is supported by the F-tests between Model E and Model F, which have a level of confidence of 3.75$\sigma$ for the ``2000,2005'' spectrum and 4.19$\sigma$ for the ``Total'' spectrum (Table \ref{tab:ftest_by_time}). The Doppler broadening component introduced by \textit{rdblur} was able to fully capture the red wing of the line, while the \textit{mytorus} model alone was not. This strongly indicates that the asymmetry in the Fe K$\alpha$ line is due to Doppler broadening and not to Compton scattering, as both Model D and Model E were unable to fit the red wing in the line, but when introducing the \textit{rdblur} component in Model F, the model was able to fully capture the asymmetry.

\begin{figure*}
\centering
\textbf{Figure 1. Spectral Fits with Gaussian Models}
{\includegraphics[scale=0.25]{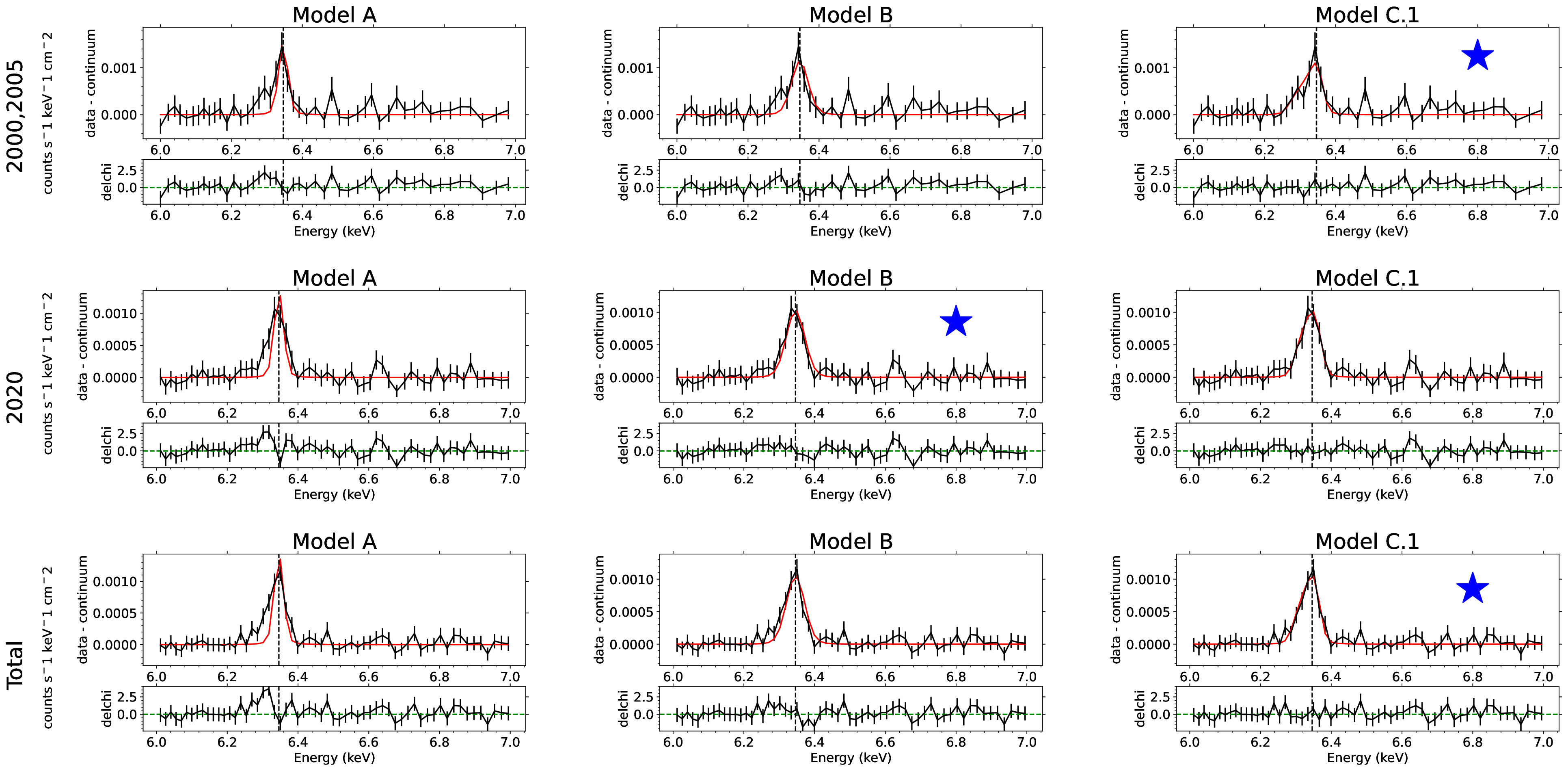}}
\caption{Fits to the Fe K$\alpha$ line using three Gaussian models: a narrow Gaussian (Model A, column 1), a broad Gaussian (Model B, column 2), and a broad Gaussian with relativistic Doppler broadening with line emissivity \textit{q} fixed to 3 (Model C.1, column 3). Model C.2, a broad Gaussian with relativistic broadening with line emissivity \textit{q} free to vary, is not shown. Row 1 is the ``2000,2005'' spectrum, row 2 is the ``2020'' spectrum, and row 3 is the ``Total'' spectrum. Below each plot is a plot of delchi versus energy; delchi uses XSPEC's definition of (data - model) / error. In these plots, ``data'' is the continuum-subtracted spectrum. A dotted line at 6.346 keV is drawn to indicate the expected energy of the peak of the line after accounting for the source redshift \textit{z} = 0.00849. The star in each row represents the benchmark model for each spectrum. The fit parameters and their errors are listed in Table \ref{tab:fits_by_time}.}
\label{fig:gaussian_fits_by_time}
\end{figure*}

\begin{figure*}
\centering
\textbf{Figure 2. Spectral Fits with Reflection Models}
{\includegraphics[scale=0.25]{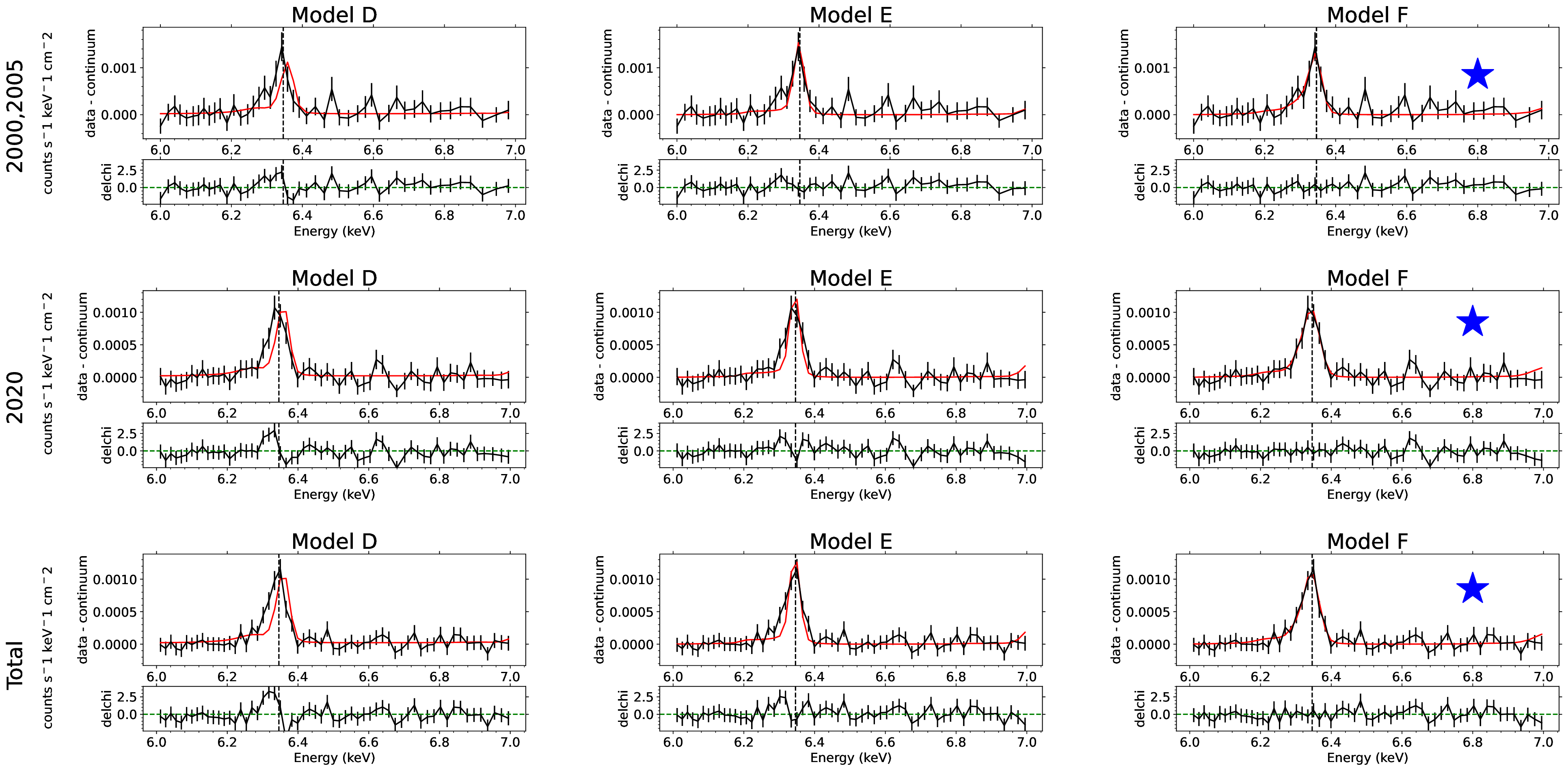}}
\caption{Fits to the Fe K$\alpha$ line using three reflection models: \textit{xillver} (Model D, column 1), \textit{mytorus} (Model E, column 2), and \textit{mytorus} with relativistic Doppler broadening (Model F, column 3). See the caption of Figure \ref{fig:gaussian_fits_by_time} for more figure details. Model D and Model E do not fully capture the red wing. Only after introducing Doppler broadening with Model F, we have fits that adequately capture the red wing. The fits with Model F compared to Models D and E demonstrate that the asymmetry in the line must be due to Doppler broadening and not Compton scattering. The star in each row represents the benchmark model for each spectrum. The fit parameters and their errors are listed in Table \ref{tab:fits_by_time}.}
\label{fig:reflection_fits_by_time}
\end{figure*}

\begin{deluxetable*}{llh||ll|lllll|ll|l}[t]
\tabletypesize{\scriptsize}
\tablecaption{Best-Fit Parameters\label{tab:fits_by_time}}
\tablehead{\colhead{Spectrum} & \colhead{Model} & \nocolhead{ObsID} & \colhead{$\Gamma$} & \colhead{N$_{PL}$ (10$^{-2}$)} & \colhead{$\sigma$ (eV)} & \colhead{A$_{Fe}$} & \colhead{Incl. (deg.)} & \colhead{q} & \colhead{R$_{in}$} &  \colhead{F$_l$ (10$^{-13}$)} & \colhead{EW$_l$ (eV)} & \colhead{$\chi^2/\upsilon$}}
\startdata
\multirow{3}{4em}{2000,2005} & Model A & 2121,6187,7240 & 1.64$\pm0.04$ & 2.30$^{+0.14}_{-0.13}$ & 0* & – & – & – & – & 5.64$\pm0.89$ & 51.0$^{+8.9}_{-7.5}$ & 0.6285\\
& Model B & 2121,6187,7240 & 1.65$\pm0.04$ & 2.33$^{+0.14}_{-0.13}$ & 19$^{+11}_{-9}$ & – & – & – & – & 7.3$^{+1.5}_{-1.4}$ & 66$^{+12}_{-10}$ & 0.6253\\
& Model C.1 & 2121,6187,7240 & 1.66$\pm0.04$ & 2.35$\pm0.14$ & 0* & – & 6.4$^{+1.5}_{-1.2}$ & 3.0* & 240$^{+110}_{-60}$ & 8.4$^{+1.3}_{-1.3}$ & 77$^{+15}_{-12}$ & 0.6192\\
& Model C.2 & 2121,6187,7240 & 1.66$\pm0.04$ & 2.34$^{+0.15}_{-0.13}$ & 0* & – & $\leq 6.01$& 2.50$^{+0.43}_{-0.29}$ & 190$^{+100}_{-130}$ & 8.2$^{+1.4}_{-1.3}$ & 75$^{+19}_{-12}$ & 0.6185 \\
& Model D & 2121,6187,7240 & $\leq 1.68$& 2.41$^{+0.11}_{-0.14}$ & – & 5.0$^{+5.0}_{-3.9}$ & $\leq 18.19$& – & – & 7.1$\pm1.2$ & 180$^{+910}_{-80}$ & 0.6370\\
& Model E & 2121,6187,7240 & 1.66$\pm0.04$ & 2.35$\pm0.14$ & – & – & 60* & – & – & 7.2$\pm1.0$ & 79 $^{+13}_{-11}$ & 0.6205 \\
& Model F & 2121,6187,7240 & 1.66$^{+0.04}_{-0.03}$ & 2.37$^{+0.17}_{-0.12}$ & – & – & $\leq22.39$ & $\leq2.65$ & $\leq5657$ & 8.6$^{+1.4}_{-1.3}$ & 96$^{+30}_{-13}$ & 0.6178 \\
\hline
\multirow{3}{4em}{2020} & Model A & 22553,24753,22554,24833,22555,24849,22556,24863,24864 & 1.69$\pm0.03$ & 2.00$\pm0.09$ & 0* & – & – & – & – & 4.93$\pm0.55$ & 55.2$^{+7.1}_{-5.4}$ & 0.6879\\
& Model B & 22553,24753,22554,24833,22555,24849,22556,24863,24864 & 1.70$\pm0.03$ & 2.04$^{+0.10}_{-0.09}$ & 22.1$^{+4.6}_{-4.1}$ & – & – & – & – & 7.27$^{+0.84}_{-0.81}$ & 81.9$^{+8.9}_{-8.3}$ & 0.6680\\
& Model C.1 & 22553,24753,22554,24833,22555,24849,22556,24863,24864 & 1.70$\pm0.03$ & 2.04$^{+0.10}_{-0.09}$ & 0* & – & 9.0$^{+3.3}_{-2.0}$ & 3.0* & 620$^{+620}_{-220}$ & 7.30$^{+0.78}_{-0.77}$ & 82.6$^{+7.8}_{-8.8}$ & 0.6653\\
& Model C.2 & 22553,24753,22554,24833,22555,24849,22556,24863,24864 & 1.70$\pm0.03$ & 2.05$^{+0.10}_{-0.09}$ & 0* & – & 9.2$^{+2.3}_{-1.8}$ & 2.38$^{+0.68}_{-0.15}$ & 190$^{+550}_{-90}$ & 7.92$\pm0.89$ & 90$^{+16}_{-9}$ & 0.6650\\
& Model D & 22553,24753,22554,24833,22555,24849,22556,24863,24864 & $\leq 1.73$& 2.11$^{+0.12}_{-0.10}$ & – & 5.0$^{+5.0}_{-3.2}$ & $\leq 18.19$& – & – & 7.14$^{+0.95}_{-0.75}$ & 81.9$^{+8.9}_{-8.3}$ & 0.6842\\
& Model E & 22553,24753,22554,24833,22555,24849,22556,24863,24864 & 1.70$\pm0.03$ & 2.05$^{+0.10}_{-0.09}$ & – & – & 60* & – & – & 6.17$\pm0.65$ & 84.3$^{+9.4}_{-8.6}$ & 0.6775 \\
& Model F & 22553,24753,22554,24833,22555,24849,22556,24863,24864 & 1.72$\pm0.02$ & 2.12$^{+0.03}_{-0.12}$ & – & – & 13.2$^{+8.7}_{-4.2}$ & 2.49$^{+0.70}_{-0.36}$ & 950$^{+950}_{-620}$ & 8.04$\pm0.82$ & 111$^{+41}_{-10}$ & 0.6649 \\
\hline
\multirow{3}{4em}{Total} & Model A & 2121,6187,7240,22553,24753,22554,24833,22555,24849,22556,24863,24864 & 1.67$\pm0.02$ & 2.11$\pm0.07$ & 0* & – & – & – & – & 5.23$^{+0.46}_{-0.45}$ & 54.1$^{+4.3}_{-4.2}$ & 0.7686\\
& Model B & 2121,6187,7240,22553,24753,22554,24833,22555,24849,22556,24863,24864 & 1.68$\pm0.02$ & 2.14$\pm0.07$ & 21.6$^{+3.8}_{-3.5}$ & – & – & – & – & 7.33$^{+0.68}_{-0.67}$ & 76.1$^{+7.8}_{-6.7}$ & 0.7505\\
& Model C.1 & 2121,6187,7240,22553,24753,22554,24833,22555,24849,22556,24863,24864 & 1.68$\pm0.02$ & 2.14$\pm0.07$ & 0* & – & 7.5$^{+1.6}_{-1.2}$ & 3.0* & 440$^{+220}_{-120}$ & 7.47$^{+0.66}_{-0.65}$ & 78.0$^{+7.0}_{-5.4}$ & 0.7447\\
& Model C.2 & 2121,6187,7240,22553,24753,22554,24833,22555,24849,22556,24863,24864 & 1.68$\pm0.02$ & 2.15$\pm0.07$ & 0* & – & 7.8$^{+2.0}_{-1.4}$ & 2.45$^{+0.30}_{-0.17}$ & 210$^{+180}_{-80}$ & 7.85$^{+0.76}_{-0.77}$ & 81.8$^{+9.4}_{-6.7}$ & 0.7436\\
& Model D & 2121,6187,7240,22553,24753,22554,24833,22555,24849,22556,24863,24864 & $\leq 1.71$& 2.22$^{+0.08}_{-0.09}$ & – & 4.9$^{+2.5}_{-2.5}$ & $\leq 18.19$& – & – & 7.15$^{+0.59}_{-0.62}$ & 210$^{+820}_{-70}$ & 0.7719\\
& Model E & 2121,6187,7240,22553,24753,22554,24833,22555,24849,22556,24863,24864 & 1.68$\pm0.02$ & 2.15$\pm0.07$ & – & – & 60* & – & – & 6.47$\pm0.53$ & 81.5$^{+6.3}_{-7.0}$ & 0.7563 \\
& Model F & 2121,6187,7240,22553,24753,22554,24833,22555,24849,22556,24863,24864 & 1.70$\pm0.02$ & 2.19$^{+0.08}_{-0.07}$ & – & – & 12.0$^{+8.2}_{-3.7}$ & 8.21$\pm0.69$ & 600$^{+9300}_{-400}$ & 8.21$\pm0.69$ & 105$^{+22}_{-10}$ & 0.7429 \\
\enddata
\tablecomments{Best fit properties of fitting the spectra from 2000+2005 (top), from 2020 (middle), and from all observations (bottom) using PyXspec. Model A represents a Gaussian with sigma fixed to 0, \emph{zphabs*(zpow + zgauss)}; Model B is the same as Model A, but with sigma able to vary freely, \emph{zphabs*(zpow + zgauss)}; Model C.1 is the same as Model B but with the Gaussian multiplied by ``rdblur'', \emph{zphabs*(zpow + rdblur*zgauss)} and line emissitivity q fixed to 3. Model C.2 is the same as Model C.1, but with line emissitivity free to vary between 2 $\leq$ q $\leq$ 4, \emph{zphabs*(zpow + rdblur*zgauss)}. Model D represents the disk reflection model xillver, which accounts for Compton scattering of the Fe K$\alpha$ line off a disk, \emph{zphabs*(zpow + xillver)}; Model E represents the reflection model mytorus, which accounts for Compton scattering off of a torus, \emph{zphabs*(zpow + mytorus)}. Model F is the same as Model E but with ``mytorus'' multiplied by ``rdblur'' and line emissivity and inclination free to vary. In all cases, the energy of the Fe K$\alpha$ line is fixed at 6.40 keV, and redshifts of all components are set to 0.00849, the redshift of MCG-5-23-16 according to \url{https://ned.ipac.caltech.edu}. The inner radius $R_{in}$ is given in units of gravitational radii $r_g = GM/c^2$, the power law normalization $N_{PL}$ in units of 10$^{-2}$ photons/keV/cm$^2$/s, and line flux $F_l$ in units of 10$^{-13}$ ergs cm$^{-2}$ s$^{-1}$. The minimum and maximum energy ranges of ``cflux'' are 6.2 keV and 6.45 keV, respectively. 1$\sigma$ errors are given in superscripts and subscripts. Greater than and less than signs indicate the 1$\sigma$ upper bound and lower bound of the parameter, respectively. Parameters marked with an asterisk were fixed at the value given.}
\end{deluxetable*}

\begin{deluxetable*}{l||ll|l|lll}
\tabletypesize{\scriptsize}
\tablecaption{F-tests between Models\label{tab:ftest_by_time}}
\tablehead{\colhead{Spectrum} & \colhead{Previous Model} & \colhead{New Model} & \colhead{Change} & \colhead{F-statistic} & \colhead{p-value} & \colhead{Confidence}}
\startdata
\multirow{2}{4em}{2000,2005} & Model A & Model B & Allowed sigma to vary & 4.99 & 0.0257 & 1.95$\sigma$ \\
& Model B & Model C.1 & Convolved Gaussian with rdblur with fixed q = 3 & 8.80 & 0.00310 & 2.74$\sigma$ \\
& Model B & Model C.2 & Convolved Gaussian with rdblur with free q & 5.39 & 0.00472 & 2.60$\sigma$ \\
& Model E & Model F & Convolved mytorus with rdblur with free q & 2.13 & 0.0956 & 1.31$\sigma$ \\
\hline
\multirow{2}{4em}{2020} & Model A & Model B & Allowed sigma to vary & 30.2 & 4.85e-08 & 5.33$\sigma$ \\
& Model B & Model C.1 & Convolved Gaussian with rdblur with fixed q = 3 & 5.07 & 0.0246 & 1.97$\sigma$ \\
& Model B & Model C.2 & Convolved Gaussian with rdblur with free q & 3.23 & 0.0399 & 1.75$\sigma$ \\
& Model E & Model F & Convolved mytorus with rdblur with free q & 7.20 & 8.78e-05 & 3.75$\sigma$ \\
\hline
\multirow{2}{4em}{Total} & Model A & Model B & Allowed sigma to vary & 31.0 & 3.18e-08 & 5.41$\sigma$ \\
& Model B & Model C.1 & Convolved Gaussian with rdblur with fixed q = 3 & 10.6 & 0.00113 & 3.05$\sigma$ \\
& Model B & Model C.2 & Convolved Gaussian with rdblur with free q & 6.79 & 0.00117 & 3.04$\sigma$ \\
& Model E & Model F & Convolved mytorus with rdblur with free q & 8.47 & 1.42e-05 & 4.19$\sigma$ \\
\hline
\enddata
\tablecomments{F-tests were performed in order to determine the statistical significance of evolving from one model to the next. See caption of Table 1 for a detailed description of each model and spectrum. F-tests with Model D were not performed because Model D gave significantly worse fits to the data than all the other models, and thus the F-test would've given an negative value for the F-statistic. Evolving from Model B to Model C.1 in the ``Total'' group gives a 3.05$\sigma$ confidence level, while the same model change for the ``2000,2005'' group gives a 2.74$\sigma$ level. This indicates that the addition of the data from the observations in 2020 to the ``2000,2005'' group \emph{just} gave enough statistical significance to reach the minimum 3$\sigma$ level needed to conclude that the Fe K$\alpha$ line in MCG-5-23-16 is asymmetric.}
\end{deluxetable*}

\vspace{-7mm}

\section{Discussion} \label{sec:discussion}
\subsection{Second-Ever Evidence for Asymmetry in the Narrow Fe K$\alpha$ Line} \label{subsec:asymmetry}
We have analyzed deep \textit{Chandra}/HETG spectra of the Type 1.9 Seyfert AGN, MCG-5-23-16. We find that the narrow Fe K$\alpha$ emission line is asymmetric, likely due to relativistic Doppler broadening. Models A and B are simple Gaussians, are physically unmotivated, and do not fully account for the red wing in all three spectra. Model D gives bad fits to the spectra, indicating that the asymmetry in the line is physical and not due to Compton scattering of the emission line photons off the disk. Model E improves the fits but still does not fully capture the red wing. Model F includes relativistic Doppler broadening and fits the spectra much better. Model C.1 seems to best fit the spectra, as it captures the red wing while preserving the simple disk geometry assumed by fixing the line emissivity to \textit{q} = 3. Model C.2 only improves the fits marginally but complicates the disk model by letting \textit{q} vary freely, so we consider Model C.1 as our benchmark fit rather than Model C.2. The best fit Model C.1 suggests $R \simeq$ 200-650 $r_g$, where $r_g = GM/c^2$, which suggests that the line originates from the innermost extents of the optical BLR or closer.

The Gaussian and reflection models with relativistic Doppler broadening fully fit the asymmetry in the Fe K$\alpha$ line. Without the Doppler broadening component, the Gaussian models and even the reflection models were unable to fully capture the asymmetry in the line (e.g. they underfit the red wing). This result indicates that the asymmetry in the Fe K$\alpha$ line in MCG-5-23-16 is due to solely relativistic Doppler broadening near the black hole and not due to Compton scattering.

Our results also indicate the asymmetry in the Fe K$\alpha$ line in MCG-5-23-16 is likely to be variable over time, as the ``2000,2005'' spectrum is best fit by an asymmetric line but the ``2020'' spectrum is sufficiently fit by a broad Gaussian. The background-subtracted net count rate for the ``2000,2005'' spectrum is 0.4079$\pm$0.0028 counts s$^{-1}$ and for the ``2020'' spectrum is 0.2903$\pm$0.0015 counts s$^{-1}$; the ``2000,2005'' spectrum has higher flux than the ``2020'' spectrum, so this variability in the line could also depend on source flux level as well. This line variability would be interesting to follow-up in a future observing campaign of MCG-5-23-16, e.g. with the recent launch of the XRISM satellite, an X-ray telescope with even higher energy resolution than Chandra and thus more detailed study of the line.

This narrow Fe K$\alpha$ line asymmetry has only been confirmed previously in NGC 4151, the brightest Seyfert galaxy in X-rays. Using \textit{Chandra} grating spectra, \citet{miller2018_NGC4151} found that the line asymmetry depends on source flux level, and that the emitting region could be located as close as 50-130 $r_g$ during the high flux intervals. Additional study of the Fe K$\alpha$ emitting region by \citet{zoghbi2019_NGC4151} using reverberation mapping demonstrated that the inner disk radius is smaller than the optical BLR radius, supporting the conclusion that in NGC 4151, the line originates in the inner regions.

Our results provide the second-ever evidence of Fe K$\alpha$ line asymmetry. Asymmetry in the Fe K$\alpha$ line in both NGC 4151 and in MCG-5-23-16 are dependent on flux and the inner radius are both constrained to be on the order of 100s to 1000 $r_g$. The emitting region in MCG-5-23-16 is thus likely to be smaller than the optical BLR as well, allowing it to be able to be used as a probe for the inner AGN geometry.

The identification of asymmetry in the narrow Fe K$\alpha$ line only in MCG-5-23-16 and NGC 4151 thus far raises intriguing questions regarding the prevalence of the line asymmetry across the broader population of X-ray AGN. It is possible that we have only detected this line asymmetry in MCG-5-23-16 and NGC 4151 thus far because of a selection bias: these two AGN were intentionally selected because of they are among the brightest AGN in X-rays and thus made natural choices for analyzing the narrow Fe K$\alpha$ line. As long as these two sources are not drastically different from the rest of the X-ray AGN population in metrics such as inclination or Eddington fraction, it is not out of the question that less X-ray bright AGN would also exhibited a skewed Fe K$\alpha$ line. The absence of such observations may also be attributed to low flux. Notably, in the case of NGC 4151, the narrow Fe K$\alpha$ line displayed a more pronounced red wing during high flux states compared to low flux states, emphasizing the potential correlation between the line profile and X-ray flux levels \citep{miller2018_NGC4151}. With higher spectral resolution from XRISM, it is possible that asymmetric narrow Fe K$\alpha$ lines could be unveiled even in dimmer AGN.

Another explanation could be that both MCG-5-23-16 and NGC 4151 are observed to have X-ray absorption, albeit not reaching Compton-thick extremes. This intermediate absorption level may suggest an intermediate inclination for these two AGN where the asymmetry in the narrow Fe K$\alpha$ line is most discernible. High inclinations, where the BLR obstructs the Fe K$\alpha$ emitting region, or low inclinations, where the line-of-sight does not pass through enough of the redshifted material might hinder clear detection of the Fe K$\alpha$ line asymmetry. It is possible that other X-ray AGN possess inclinations closer to either extreme, and thus we have not been able to clearly detect the line asymmetry in other sources.

\subsection{Disk Geometry} \label{subsec:disk_geometry}
In the best fits to the spectra, a low inclination is strongly preferred in all cases: the best fit Model C.1 to the ``2000,2005'' spectrum gives \textit{i} = 6.4$^{+1.5}_{-1.2}$ degrees, \textit{i} = 9.2$^{+2.3}_{-1.8}$ degrees for the ``2020'' spectrum, and \textit{i} = 7.8$^{+2.0}_{-1.4}$ degrees for the ``Total'' spectrum (Table \ref{tab:fits_by_time}). These inclinations are consistent with those of the emitting region in NGC 4151, the only other source confirmed to have an asymmetric narrow Fe K$\alpha$ line: the best fit inclinations for NGC 4151 are $\lesssim$ 10$^\circ$ \citep{miller2018_NGC4151}.

We would like to highlight that these inclinations for the narrow component of the Fe K$\alpha$ line are much less than the inclinations of the broad component of the line in MCG-5-23-16: for the broad component, \citet{reeves2007suzaku} and \citet{zoghbi2017mcg_nustar} cite an inclination of i $\sim$ 50$^\circ$. The low inclination is also not consistent with the classification of the source as a Seyfert 1.9-AGN, since in the classical Seyfert picture, Seyfert 1.9-AGN are high inclination sources. One possible explanation for the discrepancy is since the narrow component originates from the intermediate disk or outer disk and the broad component from the inner disk, the difference in inclination hints that this source contains a possible warp or a disk whose scale height increases from the inner regions to the outer regions (e.g. a flared disk), with the viewing angle towards the source at roughly 50$^\circ$.

We also note that the Fe K$\alpha$ line flux is consistent between the ``2000,2005'' spectrum and the ``2020'' spectrum across all models A-F, which indicates a stable AGN environment around the emitting region between 2000 and 2020, such as a a stable X-ray source and/or consistent geometry surrounding the emitting region (Table \ref{tab:fits_by_time}).

\section{Conclusions} \label{sec:conclusions}
Our results provide the \textit{second-ever} evidence of asymmetry in the \textit{narrow} Fe K$\alpha$ line. The only previous confirmed detection of asymmetry in the narrow component of this line is in the source NGC 4151 \citep{miller2018_NGC4151}, the brightest Seyfert galaxy in X-rays. As one of the next brightest Seyfert galaxies in X-rays, MCG-5-23-16 serves as a prime candidate for verifying that the asymmetry in the line in NGC 4151 may be able to be generalized to other X-ray bright AGNs. In the best fit models to the Fe K$\alpha$ line, the inner radius of the line emitting region in both NGC 4151 and in MCG-5-23-16 is constrained to be on the order of R $\simeq$ 100s to 1000 $r_g$, making the narrow Fe K$\alpha$ line originate in the innermost extents of the optical BLR, the X-ray BLR. This paper in tandem with \citet{miller2018_NGC4151} illustrates a novel method for studying the circumnuclear environment around AGN in an unprecedented manner and that the narrow Fe K$\alpha$ line can be asymmetric in AGN.

Interestingly, we find that the asymmetry in the line is due to relativistic Doppler broadening and not Compton scattering, as both the Gaussian models and reflection models only fully fit the Fe K$\alpha$ line for all three spectra when Doppler broadening was included in the models. We also see evidence that the line emitting region in MCG-5-23-16 is variable over time. We detect asymmetry on the 2000 and 2005 observations but not in the 2020 data, where the flux dropped by a factor of $\sim$1.4. This indicates that the line and emitting region are changing over time. Follow-up observations of MCG-5-23-16 are necessary to examine how the line may change on different timescales.

It also does not escape our attention that there may be an additional emission line at $\sim$6.48 keV in the ``2000,2005'' and ``Total'' spectra and a weak absorption feature at $\sim$6.68 keV in the ``2020'' and ``Total'' spectra (Fig. \ref{fig:gaussian_fits_by_time}).

Our main results are as follows:
\begin{enumerate}
    \item In the AGN MCG-5-23-16, we have found the second-ever evidence of asymmetry in the narrow Fe K$\alpha$ line. The only previously confirmed detection of asymmetry in the narrow component of the line is in the source NGC 4151.
    \item Interestingly, this asymmetry is due to solely relativistic Doppler broadening near the black hole and not Compton scattering.
    \item With a radius of R $\simeq$ 200-650 \textit{r$_{g}$}, the narrow Fe K$\alpha$ line emitting region in MCG-5-23-16 is located in the innermost extents of the optical BLR, or closer.
    \item The asymmetry in the line is variable over time, indicating that the distance of line emitting region from the black hole may be changing.
    \item The narrow component of the line strongly prefers low inclinations of $\lesssim$ 10$^\circ$, while the broad component strong prefers inclinations of $\sim$ 50$^\circ$. This suggests that the AGN has a warp or a flared disk.
\end{enumerate}

\subsection{Future Steps} \label{subsec:future_steps}
We now have two confirmed sources of asymmetry in the narrow Fe K$\alpha$ line in two of the brightest AGN in X-rays, NGC 4151 and MCG-5-23-16. Future steps to study the asymmetry in the line include, but are not limited to:
\begin{enumerate}
    \item Studying the line shape in other X-ray bright sources.
    \item Studying the variability of the line using NICER.
    \item Studying the detailed shape of the line in MCG-5-23-16 and other sources using XRISM.
\end{enumerate}

\begin{acknowledgments}
This work has been supported by the CRESST II program at the NASA Goddard Space Flight Center. The material is based upon work supported by NASA under award numbers 80GSFC21M0002 and GO0-21086A. The research in this article has made use of data obtained from the \textit{Chandra} Data Archive, and the software CIAO provided by the \textit{Chandra} X-ray Center (CXC) \citep{fruscione2006ciao} and Python package PyXspec \citep{craig2021pyxspec} and HEAsoft environment \citep{heasarc2014} developed by the HEASARC Software Development team at NASA/GSFC.
\end{acknowledgments}

\bibliography{mcg-5-23-16}{}
\bibliographystyle{aasjournal}

\end{document}